\documentclass{PoS}

\def\hhref#1{\href{http://arxiv.org/abs/#1}{#1}} 
\def\PAMELA{{\sc Pamela}}
\def\AMS{{\sc Ams-02}}

\newcommand{\beq}{\begin{equation}}
\newcommand{\eeq}{\end{equation}}

\newcommand{\ie}{i.e.}

\title{A fussy revisitation of antiprotons as a tool for Dark Matter searches}

\ShortTitle{Cosmic ray antiprotons}

\author{Mathieu Boudaud \\
LAPTh, Universit\'e Savoie Mont Blanc \& CNRS, 9 Chemin de Bellevue, B.P.110 Annecy-le-Vieux, F-74941, France\\
E-mail: \email{mathieu.boudaud@lapth.cnrs.fr}}
     
\abstract{Antiprotons are regarded as a powerful probe for Dark Matter (DM) indirect detection and indeed current data from \PAMELA\ have been shown to lead to stringent constraints. However, in order to exploit their constraining/discovery power properly, great attention must be put into effects (linked to their propagation in the Galaxy) which may be perceived as subleading but actually prove to be quite relevant. We revisit the computation of the astrophysical background and of the DM antiproton fluxes fully including the effects of: diffusive reacceleration, energy losses including tertiary component and solar modulation (in a force field approximation).
Using the updated proton and helium fluxes just released by the  \AMS\  experiment we reevaluate the secondary astrophysical antiproton to proton ratio and its uncertainties, and compare it with the ratio preliminarly reported by \AMS. We find no unambiguous evidence for a significant excess with respect to expectations. Yet, some preference for a flatter energy dependence of the diffusion coefficient  (with respect to the {\sc Med} benchmark often used in the literature) starts to emerge.  Finally, we provide a first assessment of  the room left for exotic components such as Galactic Dark Matter annihilation, deriving new stringent constraints.
}

\FullConference{The 34th International Cosmic Ray Conference,\\
		30 July- 6 August, 2015\\
		The Hague, The Netherlands}

\begin{document}

\section{Introduction}

The evidence for Dark Matter (DM) comes nowadays from a number of different astrophysical and cosmological probes, sensitive to its gravitational effects. However, we are still eagerly awaiting an explicit manifestation of DM. A possibility would be to detect anomalous fluxes of cosmic rays (charged antimatter, photons, neutrinos\ldots), which is the aim of the so-called Indirect Detection strategy. Such anomalous fluxes could indeed originate from DM pair annihilations or decays in the Milky Way halo, subsequently propagated to us within the Galactic environment. In particular, antiprotons ($\bar{p}$'s) are a sensitive probe for DM~\cite{pbar_history,pbar_history2,Cirelli:2013hv,Bringmann:2014lpa,hooperon,Hooper:2014ysa}. An intrinsic reason is that the production of $\bar{p}$'s is rather universal in DM models: as long as DM particles annihilate or decay into quarks or gauge bosons, $\bar p$ copiously emerge from the hadronization process. Other reasons are that the determination of the astrophysical background is relatively under control (at least if compared to other channels) and that the Galactic propagation of $\bar{p}$'s can be better modeled than the one of other charged particles.
The \AMS\ Collaboration has now presented its preliminary measurements of the $\bar p/p$ ratio~\cite{AMS2015}, with an improved statistical precision and energy range extending to $450\,$GeV.
In this context, it is clearly timely to refine previous predictions of $\bar{p}$ production from astrophysics and from DM, in order to obtain fluxes as accurate as possible to be compared with the precise data. This is what we describe here, based on~\cite{Boudaud:2014qra,Giesen:2015ufa}. In particular, we upgrade previous computations by incorporating energy losses and diffusive reacceleration.
In addition, \AMS\ has published the measurement of the proton ($p$) spectrum~\cite{Aguilar:2015ooa} and presented the measurement of the helium ($\alpha$) one~\cite{AMS2015}. This is important for our purposes since the $p$ and $\alpha$ spectra are crucial input ingredients in the computation of the secondary $\bar{p}$ flux, which is the minimal astrophysical $\bar{p}$ background for indirect DM searches.
However, the reach of any search for exotic physics is limited by  the astrophysical uncertainties affecting the production and the propagation processes of cosmic $\bar{p}$'s in the Galaxy and in the solar system. Indeed, while the basic processes involved in the production and propagation of CR $\bar{p}$'s are rather well understood, the detailed parameters entering in such processes are far from being well determined. The $\bar p$ production, propagation and Solar modulation uncertainties can have a large impact on both the astrophysical and (in particular) the DM signal.
%
%
%
We present the state-of-the-art astrophysical $\bar{p}$ background, carefully appraising the related uncertainties.
Then, taking into account such uncertainties, we assess what can be said on the room left for a DM signal, and what can not.


\section{Antiproton propagation in the Galaxy}
\label{propagation}

Antiprotons are deflected by the Galactic magnetic field and their transport may be seen as a diffusion process where the irregularities of this turbulent field play the role of scattering centers. In full generality, the master equation for the energy and space distribution function $f = dn/dT$ of charged cosmic rays can be written as
\begin{equation}
\frac{\partial f}{\partial t} - K(T) \! \cdot \! \nabla^2f +
\frac{\partial}{\partial z}\left\{ {\rm sign}(z) \, f \, V_{\rm conv} \right\} +
\frac{\partial}{\partial T}\left\{ b(T , \vec{x}) f - K_{\rm EE}(T) \frac{\partial f}{\partial T} \right\} = Q.
\label{eq:transport}
\end{equation}
The first term on the left hand side (l.h.s) is set to zero since one is interested in steady state conditions.
The second term accounts for space diffusion and, for $\bar{p}$'s, can be simply modeled as $K(T) = K_0 \, \beta \, ({q}/{\rm GeV})^\delta$ in terms of the kinetic energy $T$ and the momentum $q$ of the particle. The parameters $K_0$ and $\delta$ set the normalization and momentum dependence.
The third term corresponds to the convective processes, with characteristic velocity $V_{\rm conv}$, which originate in the disk and tend to push vertically outwards (hence the $z$ gradient) the $\bar{p}$ fluxes. The resulting Galactic wind reaches its nominal value of $\pm V_{\rm conv}$ right outside the disk, assumed here to be infinitely thin.
The fourth term (inside the energy derivative) accounts for energy losses, which are in general energy and space dependent. The thin disk approximation leads to express $b(K , \vec{x})$ as $2 h \delta(z) \, b(T)$, where $h = 100$~pc is the half-height of the disk. Three processes of energy loss are encoded in the negative coefficient $b$. First, $\bar{p}$'s undergo ionization losses in the interstellar neutral matter.
Then, Coulomb energy losses take place on the fraction of the interstellar medium (ISM) that is completely ionized. That mechanism is dominated by scatterings on thermal electrons. Finally, convective processes also induce a loss of energy through the conservation of the CR density in phase-space.
The last term on the l.h.s. represents diffusive reacceleration. This mechanism is produced by the drift with velocity $v_{a}$ of the diffusion centers, {\ie} the knots of the turbulent Galactic magnetic field. This yields a second order Fermi acceleration which boils down to a diffusion in energy space.
On the right hand side, the equation features the source term $Q$, which can contain different contributions.
The spallation of high-energy cosmic rays on the interstellar gas produces $\bar{p}$'s (so called `secondary') which are the source of the astrophysical background. The annihilations or decays of DM produce (so called `primary') $\bar{p}$'s.
$Q$ contains also a sink term, due to the annihilations of the $\bar{p}$'s on the interstellar gas.
Last but not least, $Q$ contains a source term (or rather `recycling' term) corresponding to tertiary $\bar{p}$'s. This term identifies the particles emerging from inelastic and non-annihilating interactions of primary or secondary
$\bar{p}$'s on the ISM. A $\bar{p}$ can collide on a proton at rest and transfer enough energy to excite it as a $\Delta$ resonance. The $\bar p$ typically loses a fraction of its energy and is effectively reinjected in the flux with a degraded momentum, flattening their spectrum as first remarked in \cite{ber99}.
In order to solve the transport equation~(\ref{eq:transport}), we model the magnetic halo of the Milky Way by a flat cylinder with half-height $L$ and radius $R$~=~20~kpc, inside which cosmic rays diffuse. The Galactic disk lies in the middle at $z=0$ and is assumed to be infinitely thin as discussed above. The CR densities $f \equiv {dn}/{dT}$ are assumed to be axi-symmetric. 
The transport equation is solved using Bessel expansion and the $\bar{p}$ flux at the Earth is derived as explained in~\cite{Donato:2001ms,Bringmann:2006im}.
Energy losses (including the tertiary component) and diffusive reacceleration (denoted 'ELDR' in the following) are generally neglected for $\bar{p}$ flux calculation. We nevertheless fully included them in our detailed analyses.
Finally, $\bar p$'s have to penetrate into the heliosphere, where they are subject to the phenomenon of Solar modulation (SMod). We describe this process in the usual force field approximation~\cite{Gleeson:1968zza}, parameterized by the Fisk potential $\phi_F$, expressed in GV.
%
%
\section{ELDR effects on the astrophysical and DM antiproton fluxes}
\label{Section_2}

\begin{figure}[!h]
\begin{center}
\includegraphics[width=0.425\textwidth]{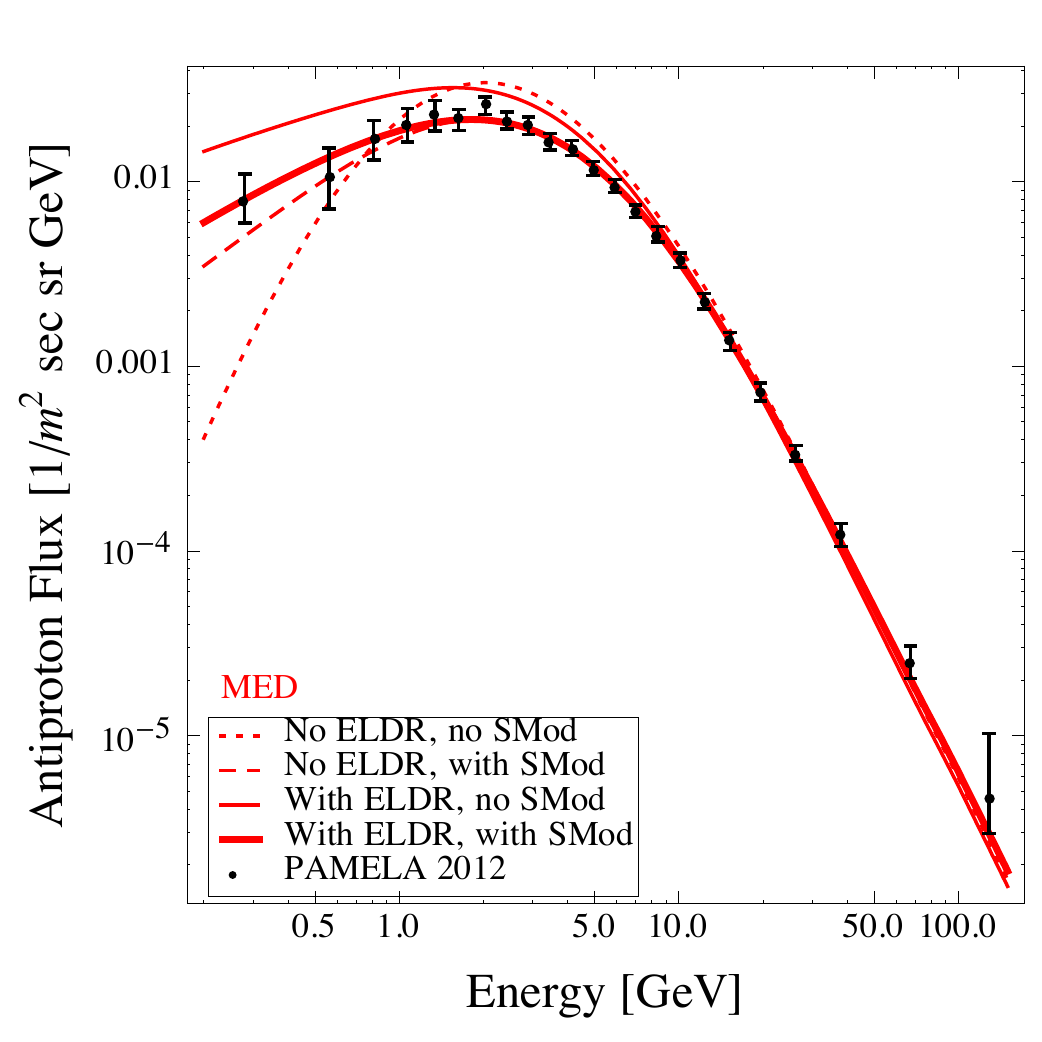} 
\includegraphics[width=0.45\textwidth]{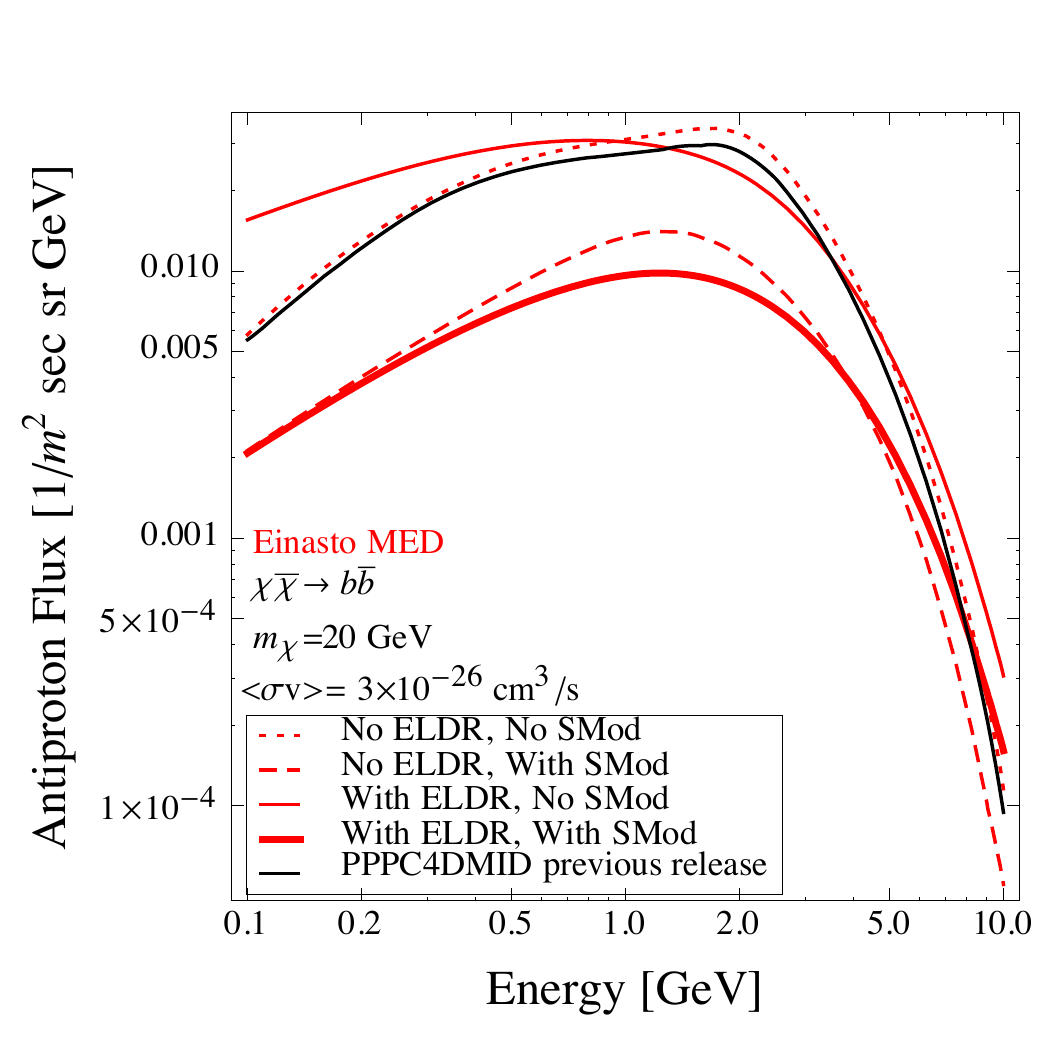}
\caption{\small \em {\bf {\em Left}}: The astrophysical background (secondary) $\bar{p}$ spectra with/without SMod and with/without ELDR,
for the MED propagation model. $\phi_F$ is set to {\rm 0.70~GV} with ELDR and {\rm 0.74~GV} without. Superimposed are the data points from \PAMELA. {\bf {\em Right}}: The $\bar{p}$ flux from Dark Matter annihilating into $b\bar{b}$ for $m_\chi = 20 \, \rm{GeV}$, $\langle \sigma v \rangle = 3 \times 10^{-26} \rm cm^{3}s^{-1}$ and an Einasto profile with/without SMod, with/without ELDR and with the previous {\sc Pppc4dmid} flux (black), for the MED propagation model. $\phi_F$ is arbitrarily set to {\rm 0.70~GV}.
}
\label{fig:ELDR_effects}
\end{center}
\end{figure}

The astrophysical $\bar{p}$ background is produced by the collisions of high-energy $p$ and $\alpha$ on the ISM, which is assumed here to be mostly composed of hydrogen and helium.
We use the injection $p$ and $\alpha$ fluxes at the Earth as measured by the \PAMELA\ experiment~\cite{pamela_p_he_2011}. Following previous studies~\cite{Donato:2001ms,Bringmann:2006im}, the Bessel transforms of these fluxes are calculated for each CR propagation model in order to derive the proton and helium densities $f_{p}$ and $f_{\alpha}$ all over the Galactic magnetic halo. The radial profile of the sources of primary CR nuclei comes into play and can be determined from pulsar and surpernova remnant surveys. We have used here the parameterization of~\cite{yusifov_kucuk}, slightly modified by~\cite{Bernard:2012wt}. In the case of a $p$ impinging on a hydrogen atom at rest, the production rate peaks around a few GeV and the involved cross-section is borrowed from~\cite{diMauro:2014zea}.
$\bar{p}$'s can also be produced in reactions involving helium nuclei either in the cosmic radiation or in the ISM. We have used the procedure discussed in~\cite{Bringmann:2006im} to which we refer the reader for details.
In the left panel of Fig.~\ref{fig:ELDR_effects}, we plot the fluxes that we obtain for the MED model of propagation~\cite{Donato:2003xg} in order to compare qualitatively the astrophysical $\bar p$ fluxes, including or not ELDR and SMod effects, with Pamela 2012 data~\cite{Adriani:2012paa}.
We show in~\cite{Boudaud:2014qra} that adding ELDR has a limited impact above 10 GeV. However, when the whole spectrum is considered, then ELDR allows a significantly better agreement with the data.

Primary $\bar{p}$'s originate from DM annihilations in each point of the galactic halo. Hence they constitute, for the purpose of the transport equation (\ref{eq:transport}), a source term $Q$ which reads 
\beq 
Q^{\rm prim}_{\bar p} = \frac{1}{2} \left(\frac{\rho}{M_{\rm DM}}\right)^2 f^{\rm ann}_{\rm inj},\qquad f^{\rm ann}_{\rm inj} = \sum_{f} \langle \sigma v\rangle_f \frac{dN_{\bar p}^f}{dK} \qquad.
\label{eq:Qann}
\eeq
The above formula shows the well known factorization of the source term in a portion that depends essentially on astrophysics (the DM density distribution $\rho$) and in a portion ($f_{\rm inj}^{\rm ann}$) that depends on the particle physics model. Here $dN_{\bar p}/dT$ are the $\bar{p}$ spectra per single annihilation or decay event and $f$ runs over all the channels with $\bar p$ in the final state, with the respective thermal averaged cross sections $\sigma v$. For additional details, we refer to~\cite{PPPC4DMID}.
The right panel of Fig.~\ref{fig:ELDR_effects} presents the effects of ELDR and SMod on the DM $\bar{p}$ spectra for $m_\chi = 20 \, \rm{GeV}$ and $\langle \sigma v \rangle = 3 \times 10^{-26} \rm cm^{3}s^{-1}$ in the case of the MED model.
Including ELDR, the peak decreases, the low energy tail is softened and the high energy portion can be somewhat raised.

\vspace{-0.7cm}
%
%



\section{Updated astrophysical background with AMS-02 data and uncertainties}
\label{Section_3}

The main ingredients of the computation for the secondary astrophysical $\bar p$ source term are: i) the injection $p$ and $\alpha$ primary fluxes from Galactic sources, ii) the collision cross sections, iii) the propagation details.
For the $p$ and $\alpha$ spectra needed in i), we now use the data that have just been released by \AMS ~\cite{Aguilar:2015ooa,AMS2015}. The spectra are measured up to a rigidity of 1.8 and 3 TV for $p$ and $\alpha$ nuclei, respectively, and they cannot be described by a single power law: a hardening at energies higher than $\sim$300 GV is observed for both.
In practice, we perform our own fits of the \AMS\ data points. The value of the Fisk potential  which gives the best $\chi^2$ for our fits is $\phi_F=0.62 \rm~GV$, the upper bound of the interval set in \cite{Aguilar:2015ooa}. The uncertainties on the slope of the $p$ and $\alpha$ spectra  at high energies, $\Delta\gamma_{p,\alpha}$, induce the purple uncertainty band on the predicted astrophysical $\bar p/ p$ ratio of Fig.~\ref{fig:rasta_constraints}.
For the production processes we need the cross sections $\sigma_{p{\rm H}\to\bar p {\rm X}}$, $\sigma_{p{\rm He}\to\bar p {\rm X}}$, $\sigma_{{\rm \alpha H}\to\bar p {\rm X}}$, $\sigma_{{\rm \alpha He}\to\bar p {\rm X}}$, where the first index refers to the impinger primary CR while the second one to the target interstellar material. For $\sigma_{p{\rm H}}$, we still use the parameterization recently proposed in~\cite{diMauro:2014zea}.
On the basis of isospin symmetry, one would consider the production cross section for antineutrons (e.g.~$\sigma_{p{\rm H}\to\bar n {\rm X}}$ and the others) as equal to those for $\bar{p}$'s; the antineutrons then rapidly decay and provide an exact factor of 2 in the $\bar p$ flux. However, as pointed out in~\cite{diMauro:2014zea,Kappl:2014hha} and as already implemented in~\cite{Boudaud:2014qra}, it may be that this na\"ive scaling does not apply and that the antineutron cross section is larger by up to 50\% with respect to the $\bar p$ one.
All these cumulated effects contribute to the red uncertainty band for the astrophysical $\bar p/ p$ ratio in the left panel of Fig.~\ref{fig:rasta_constraints}.
Once produced, $\bar{p}$'s have to propagate in the local Galactic environment before they are collected at the Earth.
The propagation parameters governing diffusion and convection are as usual codified in the {\sc Min}, {\sc Med} and {\sc Max} sets~\cite{Donato:2003xg}. Note that these have not (yet) been revised in the light of recent secondary data such as the preliminary B/C ratio of \AMS, thus the viability of  these predictions for the $\bar p/p$ ratio (which extends for instance to higher energies) is not trivially expected to hold. In left panel of Fig.~\ref{fig:rasta_constraints}, the yellow band shows the impact of the propagation uncertainty.
Finally, the value taken by $\phi_F$ is uncertain, as it depends on several complex parameters of the Solar activity and therefore ultimately on the epoch of observation. In order to be conservative, we let $\phi_F$ vary in a wide interval roughly centered around the value of the fixed Fisk potential for protons $\phi^p_F$  (analogously to what was done in~\cite{Cirelli:2014lwa}, approach `B'). Namely, $\phi_F = [0.3, 1.0]\ {\rm GV} \simeq \phi^p_F \pm 50\% \, {\phi^p_F }$. The green band in the left panel of Fig.~\ref{fig:rasta_constraints} shows this uncertainty.
We add in the left panel of Fig.~\ref{fig:rasta_constraints} the new (preliminary) \AMS\ data.
The crucial observation is that the astrophysical flux, with its cumulated uncertainties, can reasonably well explain the new datapoints. Thus, our first ---and arguably most important--- conclusion is that, contrary to the leptonic case, {\it there is no clear antiproton excess that can be identified in the first place, and thus, at this stage, no real need for primary sources}. This also means that, at least qualitatively, one expects a limited room left for exotic components, such as DM.
In order to compute constraints on DM properties, we have to identify specific sets of astrophysical parameters to describe the background. We fix in turn {\sc Min}, {\sc Med} and  {\sc Max} and we vary the Solar modulation potential in the given interval. We model the uncertainties of the production cross sections term by allowing a renormalization of the background with an energy dependence and an amplitude $A$. With this strategy, we look for the best fitting values of the amplitude $A$  and of the potential $\phi_F$. 
We find that the {\sc Min} propagation scheme predicts an astrophysical background that can {\em not} reproduce the new $\bar p/p$ data points above 30 GeV. The {\sc Med} scheme provides a barely decent fit (still good up to $\sim$ 30 GeV but rapidly degrading after) while choosing {\sc Max} the data can be well explained across the whole range of energies.
This is our second conclusion: {\em the preliminary $\bar p/p$ \AMS\ data seem to prefer a model, such as {\sc Max}, characterized by a relatively mild energy dependence of the diffusion coefficient at high energies}.


\section{Constraints on dark matter}

\begin{figure}[h!]
\begin{center}
\includegraphics[width=0.496\textwidth]{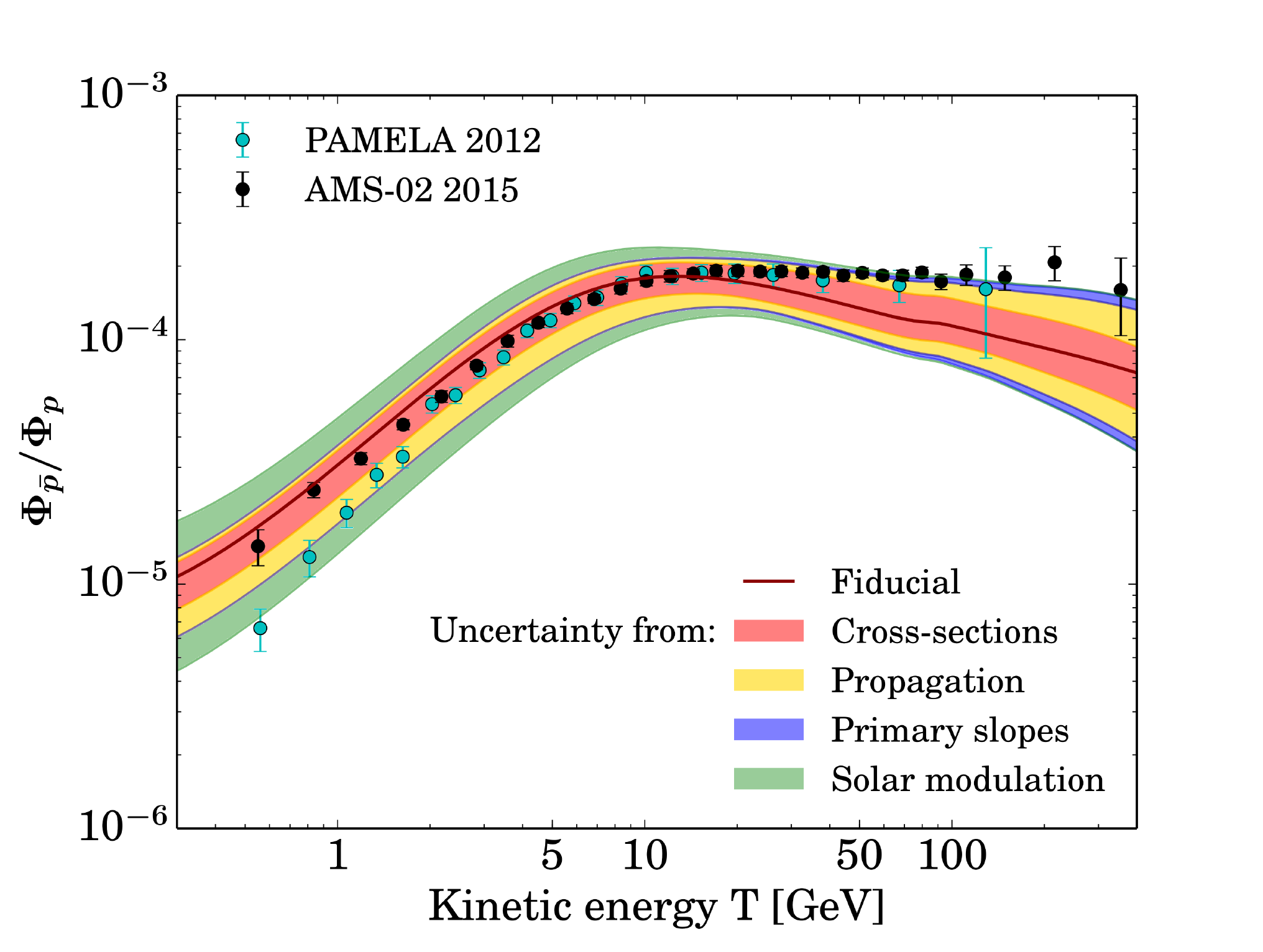}
\includegraphics[width=0.496\textwidth]{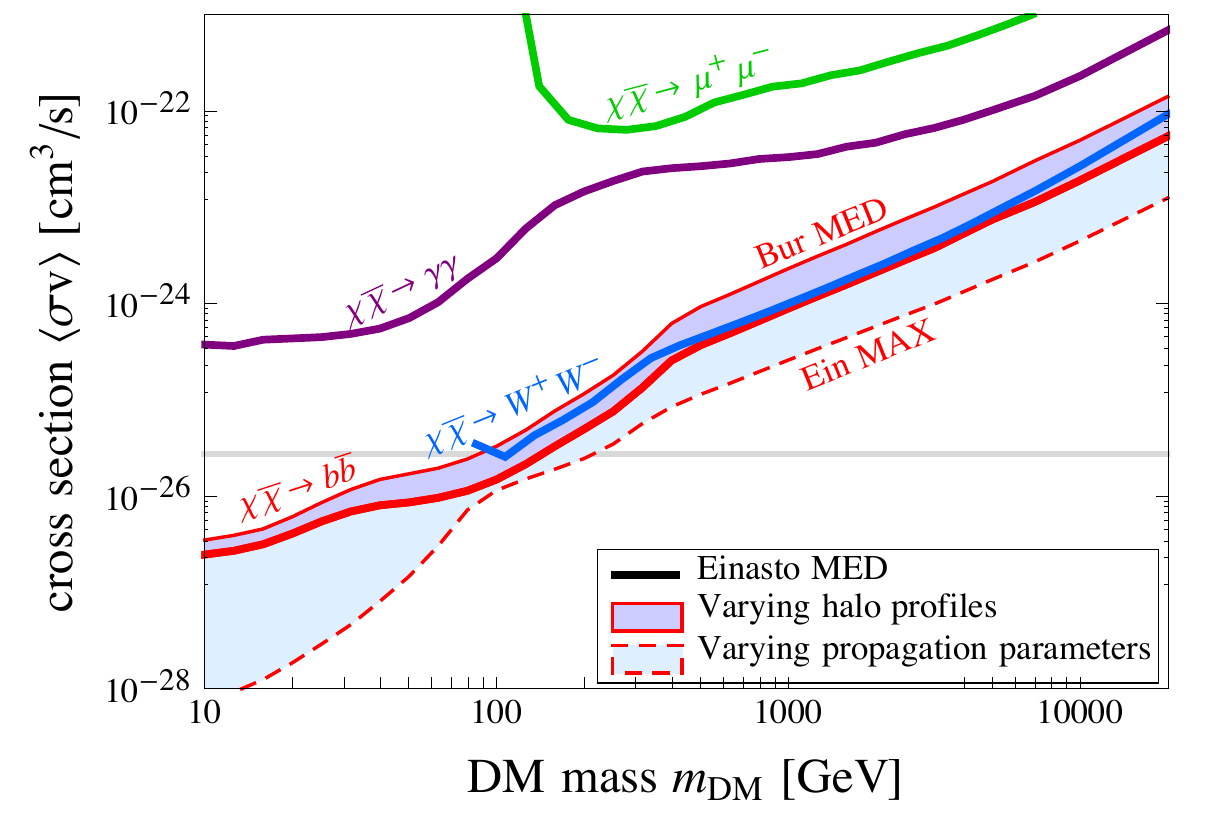}
\caption{\small \em {{\bf {\em Left}}: The combined total uncertainty on the predicted secondary $\bar p/p$ ratio, superimposed to the older \PAMELA\ data~\cite{Adriani:2012paa} and the new \AMS\ data. The curve labelled `fiducial' assumes the reference values for the different contributions to the uncertainties: best fit $p$ and $\alpha$ fluxes, central values for the cross sections, {\sc Med} propagation and central value for the Fisk potential. We stress however that the whole uncertainty band can be spanned within the errors. {\bf {\em Right}}: Estimated constraints on annihilating DM from the $\bar p/p$ ratio measurements by \AMS\ for different annihilation channels.} The areas above the curves are excluded. An illustration of the impact of DM-related astrophysical uncertainties on the constraint for the $b \bar b$ channel is given by the span of the shaded band when varying the propagation parameters (dashed line) or the halo profile (solid line).}
\label{fig:rasta_constraints}
\end{center}
\end{figure}

We can now compute the constraints in the usual plane `mass $m_{\rm DM}$ vs. thermally averaged annihilation cross section $\langle \sigma v \rangle$'. We refer to~\cite{Boudaud:2014qra} for a detailed discussion of the practical procedure.
The results that we obtain with this strategy are presented in the right panel of Fig.~\ref{fig:rasta_constraints}. We fix a benchmark DM profile (Einasto) and the {\sc Med} propagation model, and show the constraints for the different particle physics channels. We see for example that the thermal annihilation cross section $\langle \sigma v \rangle = 3 \times 10^{-26}\, {\rm cm}^3{\rm s^{-1}}$ now crosses the exclusion line for $m_{\rm DM} \sim 150$ GeV for the $\bar b b$ channel. Then, we explore the impact of changing the propagation parameters or the DM distribution. As already highlighted several times in the literature, the effect is sizable and can reach a factor of up to an order of magnitude.
Of course, as {\sc Max} maximizes by definition the DM $\bar p$ yield, its constraints are much stronger than those of the  {\sc Med} case. Turning the argument around, if the preference for {\sc Max}-like propagation schemes hinted at by preliminary \AMS\ data is confirmed, \AMS\ itself has the unprecedented capability to exclude $m_{\rm DM} \lesssim$ 250 GeV for thermal annihilation cross section in the $\bar b b$ channel.

%
%


\section*{Conclusion}

In this presentation we have reassessed the computation of the astrophysical and DM $\bar{p}$ fluxes by including effects such as Energy Losses (including tertiary component) and Diffusive Reacceleration (`ELDR'), as well as Solar Modulation (`SMod'). These effects are often perceived as subdominant, however they can actually have an important impact, especially at low energies (hence in particular for small DM masses, $\lesssim 50$ GeV).
In the light of the new $p$ flux published by \AMS\  and  the preliminary  \AMS\  results presented on the $\alpha$ flux as well as the $\bar p/p$ ratio, and using the new results of the $\bar p$ production cross sections, we have re-evaluated the secondary astrophysical predictions for the $\bar p/p$ ratio. We have accounted for the different sources of uncertainties: namely on the injection fluxes, on the production cross sections, on the propagation process and those connected to Solar modulation. Our first and main result is that {\em there is no unambiguous $\bar{p}$ excess that can be identified in the first place, and thus, at this stage, no real need for primary sources of $\bar{p}$'s}. Within errors, secondary astrophysical production alone can account for the data.
We find that the data seem to prefer a model, such as {\sc Max}, characterized by a relatively mild energy dependence of the diffusion coefficient at high energies.
We have computed the DM $\bar p$ fluxes, which we provide in the new release of P{\sc ppc}4{\sc dmid}~\cite{PPPC4DMID}.
An important application concerns updated constraints on DM: within the framework of the propagation schemes, we derive bounds that are more stringent by about one order of magnitude with respect to the previous ones~\cite{Cirelli:2013hv,Boudaud:2014qra} (based on \PAMELA\ data).
Of course, this analysis is very preliminary and there is still room for improvements. First and foremost, the release of the final $\bar p/p$ measurement with systematic and statistical errors fully accounted for. Yet, even a preliminary analysis allows to show that $\bar{p}$'s confirm themselves as a very powerful probe for CR physics and for DM in particular.
In this context, while follow-up releases of $\bar{p}$ data (e.g. pure fluxes, extended energy range or enlarged statistics) will  obviously be welcome, it is urgent to first address one of the main current limitations in the field of charged CRs, namely the determination of the propagation parameters. In this respect, analyzing the upcoming reliable and accurate light nuclei measurements from \AMS\ will provide the community with a very powerful leverage for any search of exotics in CR's. Only then will it be possible to assess whether or not excesses are present in $\bar{p}$ data, although identifying their origin will remain very challenging~\cite{Pettorino:2014sua}.





\end{document}